\documentclass[aps,preprint,amsmath,amssymb]{revtex4-1}

\usepackage{graphicx, color}

\def\beq{\begin{equation}}
\def\eeq{\end{equation}}
\def\be{\begin{eqnarray}}
\def\ed{\end{eqnarray}}
\def\non{\nonumber}
\def\ga{\gamma}

\begin{document}
\preprint{DESY 12-027}
\title{\Large \bf CP violation in $D^0 \to (K^- K^+ , \pi^- \pi^+)$ from diquarks  }

\date{\today}

\author{ \bf  Chuan-Hung Chen$^{1,2}$\footnote{Email:
physchen@mail.ncku.edu.tw}, Chao-Qiang Geng$^{3,2}$\footnote{Email: geng@phys.nthu.edu.tw} and Wei Wang$^4$\footnote{Email:wei.wang@desy.de}
 }

\affiliation{ $^{1}$Department of Physics, National Cheng-Kung
University, Tainan 701, Taiwan \\
$^{2}$National Center for Theoretical Sciences, Hsinchu 300, Taiwan\\
$^3$ Department of Physics, National Tsing Hua University, Hsinchu 300, Taiwan\\
$^4$ Deutsches Elektronen-Synchrotron DESY, Hamburg 22607, Germany
 }

\begin{abstract}
The explanation of  the  large CP asymmetries in $D^0\to (\pi^+\pi^-, K^+K^-)$ decays observed by the LHCb collaboration is likely to call for new physics   beyond the CKM paradigm.   We explore  new contributions caused by  the color-sextet scalar diquark, and demonstrate that the diquark with the mass of order 1 TeV and nominal couplings with quarks can generate  the CP asymmetries  at the percent level.  Using the experimental data on branching ratios and  CP asymmetries of  $D^0\to (\pi^+\pi^-, K^+K^-)$, we derive the constraints on the diquark mass and couplings, which can be further examined in hadron colliders in the dijet final states. 

\end{abstract}
\maketitle

It is generally anticipated that   both direct and indirect CP asymmetries (CPAs)  in the charm sector
 are quite small in the standard model (SM). Any observation of the large CPA in $D^0$ decays  will presumably imply 
 that the  underlying theory is out of the scope of  the SM. 

Recently  based on the  0.62 fb$^{-1}$ of data  collected  in 2011,  the LHCb collaboration~\cite{LHCb} has measured 
the difference between the time-integrated CP asymmetries in the
decays $D^0 \to K^+K^-$ and $D^0 \to \pi^+\pi^-$,
$\Delta A_{CP} \equiv A_{CP}(D^{0}\to K^{+}K^{-})-A_{CP}(D^{0}\to \pi^{+}\pi^{-})$, 
given by
\beq\label{LHCbData}
\Delta A_{CP} =(-0.82\pm0.21(\text{stat.})\pm0.11(\text{sys.}))\%\,,
\eeq
where  the first uncertainty is statistical and the second is systematic.  
The quantity $ A_{CP}(D^{0}\to f)$  is defined as
\beq\label{CPA}
A_{CP}(D^{0}\to f) \equiv \frac{\Gamma(D^0\to f)-\Gamma(\bar D^0\to f)}
  {\Gamma(D^0\to f)+\Gamma(\bar D^0\to f)}\,,
\eeq
with $f= K^+K^-, \pi^+\pi^-$. By contrast,    results released by the CDF
collaboration~\cite{CDF2011} based on 5.9~fb$^{-1}$ of the integrated luminosity   are somewhat less conclusive, given by
\be\label{eq:CDFData}
A_{CP}(D^{0}\to \pi^{+}\pi^{-})&=& (+0.22\pm0.24 \pm0.11) \%\,,
\non\\
A_{CP}(D^{0}\to K^{+}K^{-})&=& (-0.24\pm0.22 \pm0.09) \%\,,
\ed
while the previous world averages from Heavy Flavor Averaging Group~\cite{hfag} in 2010 are
\be\label{hfagData}
A_{CP}(D^{0}\to \pi^{+}\pi^{-})&=& (+0.22\pm0.37)\%\,,
\non\\
A_{CP}(D^{0}\to K^{+}K^{-})&=& (+0.16\pm0.23)\%\,.
\ed
The new world average for $\Delta A_{CP}$ from Eqs. (\ref{LHCbData}), (\ref{eq:CDFData}) and (\ref{hfagData})
is found to be~\cite{hfagUpdate}
\beq\label{NewData}
\Delta A_{CP} = -(0.645\pm 0.180)\%\,,
\eeq
which is about $3.6\sigma$ away from zero.

Contributions to $A_{CP}(D^{0}\to f)$ contain both direct  ($A_{CP}^{dir}(D^{0}\to f)$) and indirect 
 ($A_{CP}^{ind}(D^{0}\to f)$) parts, and from the LHCb report~\cite{LHCb} one has  
\beq
\Delta A_{CP} \simeq \Delta A_{CP}^{dir}+ (9.8\pm0.3)\% A_{CP}^{ind}\,,  \label{eq:DirCP}
\eeq
where $\Delta A_{CP}^{dir}\equiv A_{CP}^{dir}(D^{0}\to K^{+}K^{-})-A_{CP}^{dir}(D^{0}\to \pi^{+}\pi^{-})$,
and $A_{CP}^{ind}=A_{CP}^{ind}(D^{0}\to f)$, which is universal for $f=K^{+}K^{-}$ and $\pi^{+}\pi^{-}$
 and less than $0.3$\% due to the mixing parameters. 
Clearly,  the LHCb data in Eq. (\ref{LHCbData}) is dominated by the difference of the direct CP asymmetries, $\Delta A_{CP}^{dir}$.

In order to have a nonzero direct CPA, two amplitudes $A_1$ and $A_2$ with both nontrivial weak  phase difference $\theta_W$ and strong phase difference
$\delta_{S}$ are called for, leading to 
\be
\label{eq:DirCPA}
&&A_{CP}^{dir}(D^0\to f) = { |A_f|^2-|\bar{A}_f|^2\over |A_f|^2+|\bar{A}_f|^2} \non \\
&=& {-2|A_1||A_2|\sin\theta_W\sin\delta_{S}\over |A_1|^2+|A_2|^2+2|A_1||A_2|\cos\theta_W\cos\delta_{S}}\nonumber\\
&\simeq& -2r_f \sin\theta_W\sin\delta_{S},
\ed
with $r_f= |A_2|/|A_1|$. 
In the last step, a hierarchy of  $r_f\ll 1$
%amplitudes 
has been adopted. 
The SM description of   the direct CPA  for $D^0\to f$ arises from the interference between tree and penguin contributions, in which decay amplitudes take the generic expressions
\begin{eqnarray}
 A_f= V_{cq}^* V_{uq} T_{\rm SM} + V_{cb}^* V_{ub} P_{\rm SM},
\end{eqnarray}
with $q=s$ for $f=K^+K^-$ and $q=d$ for $\pi^+\pi^-$. 
%By naive calculations, 
%it could be estimated by
% \be
% A^{dir}_{CP}(D^0\to f) \sim - Im\left( \frac{V^*_{cb} V_{ub}}{V^*_{cq} V_{uq}} \right)\frac{2P'_{SM} E'_{SM} %\sin\delta_S}{ |T^{\prime }_{SM}+E'_{SM}|^2 }\,,
% \ed
%where $V_{q'q}$ (q=s, d) is the CKM matrix element, $T'_{SM}(P'_{SM})$ denotes the tree (penguin) emission %topology in the SM,  $E'_{SM}$ stands for the contributions of W-exchange topology and $\delta_S$ is the %associated CP-even phase. With the SU(3) limit,
%$Im(V^*_{cb} V_{ub}/V^*_{cq} V_{uq})\approx \pm A^2 \lambda^4 \eta$,
%$E'_{SM}\sim T'_{SM}$, and $\sin(\delta_S) \sim -O(1)$,  we could have  
% \be
% A_{CP}(K^- K^+) &\sim& -A_{CP}(\pi^- \pi^+) \non \\
% &\sim& - A^2 \lambda^4 {P'_{SM}\over T'_{SM}} > - 0.07{P'_{SM}\over T'_{SM}}\%\,.  
% \ed
Besides the hierarchy in the CKM matrix elements $V_{cq}^* V_{uq}\gg V_{cb}^* V_{ub}$, penguin amplitudes are also suppressed by loop factors. 
Even in the  limit $P_{SM}\sim T_{SM}$,  the ratio of the decay amplitudes is still very small $r_f \sim 0.0007$, leading to a tiny CPA which is far below the central value in
 Eq. (\ref{NewData}). As a result, if we took the data by the LHCb seriously, a solution to the large $\Delta A_{CP}$ would be to introduce some new CP violating mechanism beyond the CKM.
 
By neglecting the small SM penguin contributions, the decay amplitude of $D^0\to f\, (f=K^+K^-$ or $\pi^+\pi^- )$ with new physics contributions could be parametrized as
 \be\label{ANP}
 A_f = V^*_{cq} V_{uq} \left[ T'_{SM} \left( 1+ \rho e^{i\theta_W}\right) + E'_{SM} e^{i\delta_S}\right]\,, \label{eq:amp}
 \ed
where $\theta_W$ is the new weak CP phase ranging from 0 to $\pi$, while $\rho$ is associated with the new physics effect with an arbitrary sign, $i.e.$, sign$(\rho)=\pm 1$.   In the above equation, $T'_{SM} $ corresponds to the $W$ emission diagram, while $E_{SM}'$ is from the annihilation type  of the W-exchange diagram. 
We note that since the final state interactions make dominant contributions to the W-exchange diagram, without losing generality, we regard that 
the short-distance (SD) effect of new physics on such topology could be ignored. Consequently, 
in this circumstance new physics plays an important role for the emission topology. 
 From Eqs. (\ref{eq:DirCPA}) and (\ref{ANP}), we find 
 \be
 A^{dir}_{CP}(D^0\to f) = \frac{4T'_{SM} E'_{SM} \rho \sin\theta_W \sin\delta_S}{| a(\zeta) |^2 +|a(\zeta\to \zeta^*)|^2 } 
 \ed
 with $a(\zeta)= T'_{SM} \zeta + E'_{SM} e^{i\delta_S} $ and $\zeta=1 + \rho \exp(i\theta_W)$. 
 If all quantities in the SM are   under control,  $\rho$ and $\theta_W$ are the only free parameters.

Recently, a number of theoretical studies~\cite{Bigi:2011,Isidori2011,Brod:2011re,Wang:2011uu,arXiv:1111.6515,arXiv:1111.6949,Hochberg:2011ru,Pirtskhalava:2011va,Chang:2012gn,Giudice:2012qq,Bhattacharya:2012ah,Cheng:2012wr,Altmannshofer:2012ur}
 have been  performed to understand the LHCb and CDF data.
 In this brief report, we would like to use the scalar diquarks as the sources of new physics.
 Although the introduction of scalar diquarks in
the literature has its physical reason, such as solving the strong CP
problem \cite{BZ_PRL55,Barr_PRD34,BF_PRD41},
here we only explore their consequences in
$D$-meson processes. The combinations of two quarks give many possible types of diquarks, such as
$(3,1,-1/3)$, $(6,1,1/3)$, $(3,3,-1/3)$, $(3,1,-4/3)$, and $(6,1,4/3)$ under the standard group of 
$SU(3)_C\times SU(2)_L \times U(1)_Y$, among which the $(3, 1, -1/3)$ and $(6, 1, 1/3)$ diquarks are very interesting  for avoiding the strong correlation in flavor couplings and the strict constraint from the tree induced $D$-$\bar D$ mixing~\cite{Chen2009}.  In what follows we will concentrate on the color sextet $H_6=(6,1,1/3)$ to illustrate its effect on the direct CPAs in $D^0\to K^+K^-,\pi^+\pi^-$, but a  similar study could be applied to the color triplet boson. 

\begin{figure}[htbp]
\includegraphics*[width=3.5 in]{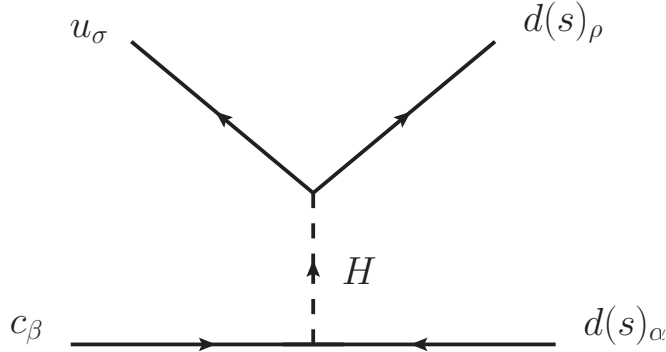}
\caption{Diquark-mediated flavor diagram for $D^0\to (\pi^+ \pi^-, K^+ K^-)$ }
 \label{fig:Dtree}
\end{figure}

We first write the interaction between quarks and $H_6$ as
 \be
{\cal L}_{H_{\bf 6}} = f_{ij} d^{c^T}_{i\alpha} C P_L u^c_{j\beta}
H^{\alpha\beta}_{\bf 6} +h.c. \,,\label{eq:LH6-1}
 \ed
where $f_{ij}$ denotes the couplings between the diquark and quark flavors,
$C=i\ga^0 \ga^2$ is the charge-conjugation matrix, $P_{L(R)}=(1\mp
\ga_5)/2$ is the chiral projection operator, and
$H^{\alpha\beta}_{\bf 6}$ is a weak gauge-singlet and colored sextet
scalar with $\alpha$ and $\beta$ being the color indices. 
In terms of the interactions in Eq.~(\ref{eq:LH6-1}), flavor
diagrams for D decays are given in Fig.~\ref{fig:Dtree}. After integrating out the highly virtual diquarks, the
corresponding interactions for $c\to u \bar d d (\bar s s)$
are derived as
 \be
 {\cal H}_{c\to u}&=& - \frac{f^*_{qc} f_{qu}}{16m^2_H}
 \left(O^{q}_{1}+O^{q}_{2} \right) \,,\non\\
 O^{q}_{1} &=& (\bar u   q)_{V+A} ( \bar q   c)_{V+A} \,,\non \\
 O^{q}_{2} &=& (\bar u_\alpha  q_\beta)_{V+A} (\bar q_\beta  c_\alpha)_{V+A}\,,
 \ed
with $(\bar q'' q')_{V+A}=\bar q'' \ga^\mu (1+\gamma_5) q'$ and $q=d, s$.
Based on the decay constants and transition form factors, defined by
 \be
\langle 0 | \bar q' \ga^\mu \ga_5 q|P(p)\rangle &=& if_P p^\mu\,, \non \\
 \langle P(p_2)| \bar q \ga_\mu c| D (p_1)\rangle &=&
f^{DP}_{+}(k^2)\Big\{Q_{\mu}-\frac{Q\cdot k }{k^2}k_{\mu} \Big\} \non \\
&+&\frac{Q\cdot k}{k^2}f^{DP}_{0}(k^2)\,k_{\mu}\,,
 \ed
respectively,
with $Q=p_1+p_2$ and $k=p_1-p_2$, the diquark contribution to $D \to \bar f$ is given by
 \be
 \rho e^{i\theta_W}&=& \frac{\sqrt{2}}{16 m^2_H G_F } \frac{f^*_{qc} f_{qu} a_1'}{V^*_{cq} V_{uq} a_1}\,, \label{eq:NPratio}
 \ed
 where $a_1 = c_1(\mu) + c_{2}(\mu) (1/N_c + \chi(\mu))$ and $a_1'= 1+(1/N_c + \chi(\mu))$ are  the effective Wilson coefficients  \cite{Cheng:2010ry}  with $\chi (\mu)$ being the nonfactorizable contribution. In the large $N_c$ limit, as the nonfactorizable effect could be simplified as $\chi=-1/N_c$ \cite{Buras:1985xv},
 %. We find that  with the limit, 
 the nonfactorizable part will be 
 smeared by the operator $O^q_2$.  
 %%%%%%%%%%%%%%%%%%%%%%%%%%%%%%%%%%%%%%%%%%%%%%%%%%%%%%%%%%%%%%%%%%%%%%%%%
%%%%%%%%%%%%%%%%%%%%%%%%%%%%%%%%%%%%%%%%%%%%%%%%%%%%%%%%%%%%%%%%%%%%%%%%%

 Before the numerical study, we discuss how to escape the  stringent constraint from the $D-\bar D$ mixing. Using the results in Ref.~\cite{Chen2009}, the $D$ mixing parameter induced by box diagrams could be formulated by 
  \be
  x_D = \frac{\Delta m_D}{\Gamma_D} \sim \tau_D \frac{19 f^2_D m_D}{1536\pi^2} \left| \frac{\sum_i f^*_{ic} f_{iu}}{m_H}\right|^2\,.
  \ed 
 Since the flavors (denoted by $i$) in the internal loops include $d$, $s$ and b quarks, the constraint from $x_D$ could be released if a cancellation occurs 
 in $\sum_i f^*_{ic} f_{iu}$.  
 
 In our analysis, the input values of the SM are taken as \cite{Cheng:2012wr,Cheng:2010ry,PDG2010}:
  \be
 && T'_{SM}(\pi \pi, K K ) = ( 3.01, 4.0 ) \times 10^{-6} \textrm{ GeV} \,, \non \\
   && E'_{SM}(\pi \pi, K K) = ( 1.3, 1.6)\times 10^{-6} \textrm{ GeV} \,, \non \\
  &&\delta_{S}( \pi \pi, K K)=( 145, 108)^{\circ}  \,,   a_1= 1.21, \non\\
  %\ f_{\pi(K)}=130 (160) \textrm{ MeV}\,,  \non \\
%    &&F^{D\pi (K)}_0= (0.666,0.739)\,,    , \non \\ 
 &&V_{us} = -V_{cd} = 0.2252\,,  V_{cs} = 0.97345,\;\; V_{ud} = 0.97428\,, \non \\
     &&  m_{\pi(K)} = 0.139 (0.497)\textrm{ GeV}\,,   m_D =1.863 \textrm { GeV}\,, 
  \ed
 where the resulting branching ratios (BRs) for $D^0\to (\pi^- \pi^+, K^-K^+)$ are estimated as $(1.38, 3.96) \times 10^{-3}$, while the current data are ${\cal B}(D^0\to \pi^- \pi^+ )=(1.400 \pm 0.026)\times 10^{-3}$ and ${\cal B}(D^0\to K^- K^+)= (3.96\pm 0.08)\times 10^{-3}$ \cite{PDG2010}.  
 Since $f^*_{dc}f_{du}$ and $f^*_{sc} f_{su}$ are independent free parameters, 
 to simplify our calculation, we adopt two benchmark schemes: (I) $f^*_{dc}f_{du} \approx f^*_{sc} f_{su}$ and (II) $f^*_{dc}f_{du}  \approx -f^*_{sc} f_{su}$. Consequently, the involved parameters in the analysis are $\theta_W=arg(f^*_{sc} f_{su})$ and $|f^*_{sc}f_{su}|/m^2_H$.

An estimate of the scalar diquark contribution is given as follows. The current measurement of the dijet cross section from the hadron collider puts the limit of the scalar diquark mass, see Ref.~\cite{MOY_PRD77,Han:2010rf} for instance, 
\begin{eqnarray}
 m_{H_6}> 1.9 {\rm TeV},
\end{eqnarray}
where a normal diquark-quark coupling is used.  
When the diquark  decay is taken into account, this value may get reduced.  Assuming the mass of order 1 TeV and normal couplings for the diquark, we find from Eq.~(\ref{eq:NPratio}) that the NP contribution  is at the percent level compared to the SM contribution.  As a result, such a diquark is able to explain the large CPA data by the LHCb, while its effects to branching ratios will be up to a few percent. 

\begin{figure}[htbp] 
\includegraphics[scale=0.8]{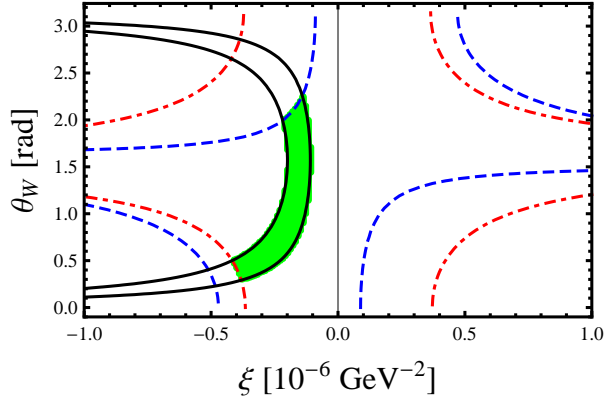} 
\caption{$\Delta A^{dir}_{CP}$ (solid) in units of $10^{-2}$ as a function of $\xi \equiv \pm |f^*_{sc} f_{su}|/m^2_H$ and $\theta_W$ in the diquark model,
where the dashed (blue) and dash-dotted lines denote the data  with errors of BRs for the decays $D^0\to (\pi^+ \pi^-, K^+ K^-)$ in units of $10^{-3}$, respectively,
while the shadowed region (green) is the combined constraint from BRs and direct CPAs.  }
 \label{fig:CP_I}
\end{figure}

Since the involved parameters in Cabibbo allowed processes, $e.g.$, $D\to \pi K$, are different from the singly Cabibbo suppressed decays, with the assumption of $\sum_i f^*_{ic} f_{iu}\to 0$, the BRs for  $D^0\to (\pi^+ \pi^-, K^+ K^- )$ are the potential constraint. Thus, we will take the data of the BRs with errors to constrain the free parameters. For Scheme I, we show the CPA difference defined in Eq.~(\ref{eq:DirCP}) as a function of $\xi \equiv \pm |f^*_{sc} f_{su}|/m^2_H$ and $\theta_W$ in Fig.~\ref{fig:CP_I}. The solid lines correspond to the upper and lower bounds of the LHCb results. The dashed (blue) and dash-dotted (red) lines denote the experimental BRs for the decays $D^0\to (\pi^+ \pi^-, K^+ K^-)$, respectively. From these curves, we obtain the allowed region of the parameters as
\begin{eqnarray}
  -2\times 10^{-7} <\xi <-0.6 \times 10^{-7},\nonumber\\
  0.3 <\theta_W <2.25.
\end{eqnarray}

%%%%%%%%%%%%%%%%%%%%%%%%%%%%%%%%%%%%%%%%%%%%%%%%%%%%%%%%%%%%%%%%%%%%%%%%%
%%%%%%%%%%%%%%%%%%%%%%%%%%%%%%%%%%%%%%%%%%%%%%%%%%%%%%%%%%%%%%%%%%%%%%%%%

\begin{figure}[htbp]
\includegraphics[scale=0.8]{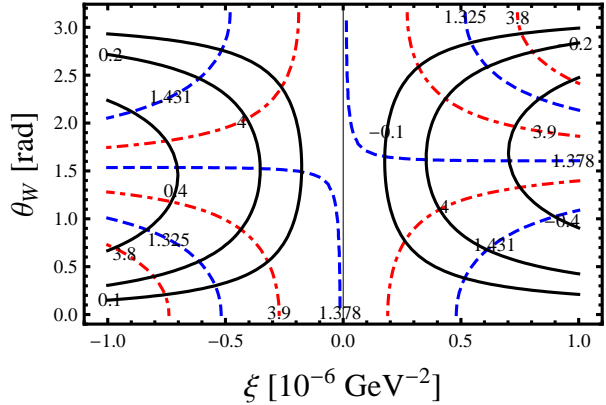} 
\caption{Legend is the same as that in Fig.~\ref{fig:CP_I}, but the magnitude of $\Delta A^{dir}_{C}$ is smaller than the LHCb results of $1\sigma$.  }
 \label{fig:CP_II}
\end{figure}

\begin{figure}[htbp]
\includegraphics[scale=0.8]{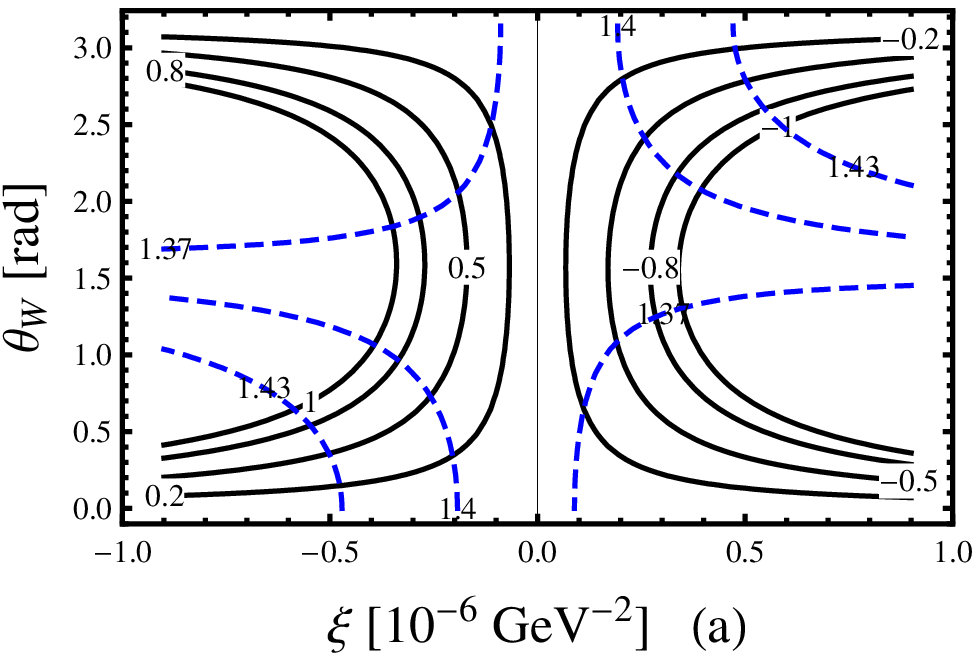}  
\includegraphics[scale=0.8]{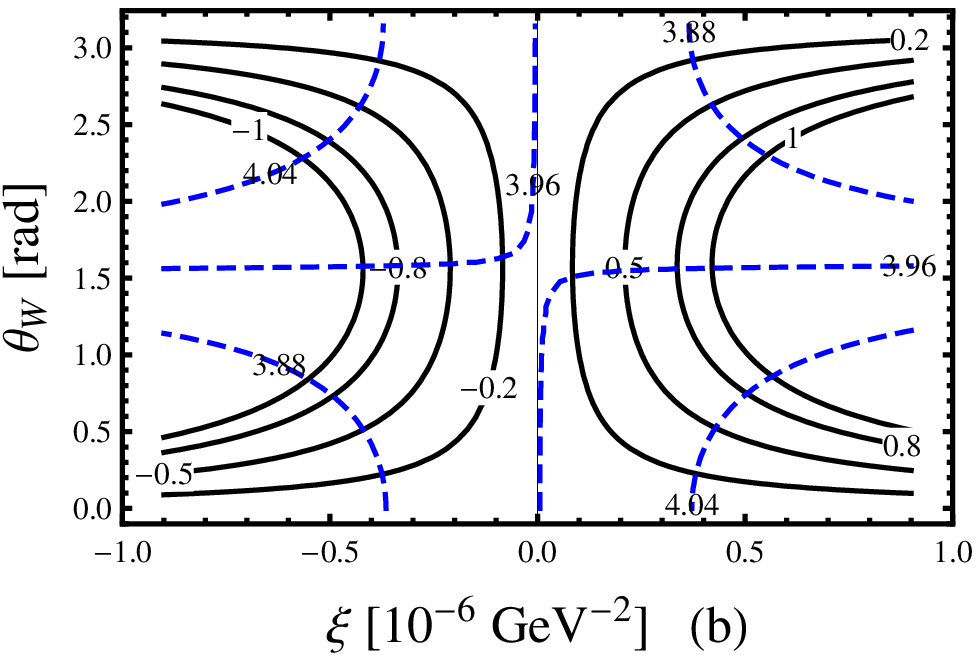} 
\caption{
% The solid line is the 
Direct CPAs (solid lines) and BRs (dashed lines)  for (a) $D^0\to \pi^+ \pi^-$ and (b) $D^0\to K^+ K^-$,
% and the dashed lines stand for the BR of the corresponding decay. T
where the units for the solid and dashed lines  are $10^{-2}$ and $10^{-3}$, respectively.
%on solid lines is $10^{-2}$ while other unit  is $10^{-3}$.  
 }
 \label{fig:CPA}
\end{figure}

Similarly, the results in Scheme II are presented in Fig.~\ref{fig:CP_II}. In this scheme, we find that the magnitude of $\Delta A^{dir}_{CP}$ cannot fit the data within $1\sigma$ errors. The reason is that  the direct CPA of $D^0 \to \pi^+ \pi^- $ in scheme II  has the same sign with the one of $D^0\to K^+K^-$. 
%%%%%%%%%%%%%%%%%%%%%%%%%%%%%%%%%%%%%%%%%%%%%%%%%%%%%%%%%%%%%%%%%%%%%%%%%
%%%%%%%%%%%%%%%%%%%%%%%%%%%%%%%%%%%%%%%%%%%%%%%%%%%%%%%%%%%%%%%%%%%%%%%%%
In order to understand the effects on each direct CPA $A^{dir}_{CP}$ defined in Eq.~(\ref{eq:DirCPA}), by using Scheme I, we display  $A^{dir}_{CP}$ for $D^0\to \pi^+ \pi^-$ and $K^+ K^-$ decays in Figs.~\ref{fig:CPA}a and \ref{fig:CPA}b, respectively. The results, even the sign, are consistent with the current CDF data shown in Eq.~(\ref{eq:CDFData}).
%%%%%%%%%%%%%%%%%%%%%%%%%%%%%%%%%%%%%%%%%%%%%%%%%%%%%%%%%%%%%%%%%%%%%%%%%
%%%%%%%%%%%%%%%%%%%%%%%%%%%%%%%%%%%%%%%%%%%%%%%%%%%%%%%%%%%%%%%%%%%%%%%%%

In summary, it is clearly an exciting moment if any CPA in  D-meson decays is observed at the percent level as the SM prediction is far below 1\%. Motivated by the
recent LHCb measurement  on the time integrated CPA in $D^0 \to (\pi^+ \pi^-, K^+ K^-)$ decays, which appears to be inconsistent with the SM result, we have studied
 the contributions of the colored scalar boson and used the color sextet $(6, 1, 1/3)$ as the illustrator. In this diqaurk model, the serious constraint from the D mixing parameter induced by box diagrams could be avoided if a cancellation  among different flavor couplings occurs. We have found that the the induced direct CPAs of  $D^0 \to (\pi^+ \pi^-, K^+ K^0)$  decays can fit the recent LHCb results and are consistent with the CDF current measurement.  \\

 \noindent {\bf Acknowledgments}

WW thanks Ahmed Ali for useful discussions. We are grateful to Prof. H.~Y.~Cheng for a communication.
This work is supported by the National Science Council of R.O.C.
under Grant \#s: NSC-100-2112-M-006-014-MY3 (CHC) and
NSC-98-2112-M-007-008-MY3 (CQG) and 
the Alexander von Humboldt Foundation (WW) .

\end{document}